\documentclass[12pt]{article}

\usepackage{amsmath}
\usepackage{amssymb}

\allowdisplaybreaks

\begin{document}

\baselineskip=18.8pt plus 0.2pt minus 0.1pt

%%%%%%%%%%% Private Macros %%%%%%%%%%%%%
\makeatletter

\@addtoreset{equation}{section}
\renewcommand{\theequation}{\thesection.\arabic{equation}}
\renewcommand{\thefootnote}{\fnsymbol{footnote}}
\newcommand{\beq}{\begin{equation}}
\newcommand{\eeq}{\end{equation}}
\newcommand{\bea}{\begin{eqnarray}}
\newcommand{\eea}{\end{eqnarray}}
\newcommand{\nn}{\nonumber\\}
\newcommand{\hs}[1]{\hspace{#1}}
\newcommand{\vs}[1]{\vspace{#1}}
\newcommand{\p}{\partial}
\newcommand{\bra}[1]{\left\langle  #1 \right\vert }
\newcommand{\ket}[1]{\left\vert #1 \right\rangle }
\newcommand{\vev}[1]{\left\langle  #1 \right\rangle }

\makeatother
%%%%%%%%% End of private macros %%%%%%%%%%%

%%%%%%%%%%%%%%%%%%%%%%%%%%%%%%%%%%%%%%%%%%%%%%%%%%%%%%%%%%%%%%%%%%%
\begin{titlepage}
\title{
\hfill\parbox{4cm}
{\normalsize\tt %arXiv:yymm.nnnn
}\\
\vspace{1cm}
On gauge transformation property of coordinate independent SO(9) vector states in SU(2) Matrix Theory
}
\author{Yoji Michishita
\thanks{
{\tt michishita@edu.kagoshima-u.ac.jp}
}
\\[7pt]
{\it Department of Physics, Faculty of Education, Kagoshima University}\\
{\it Kagoshima, 890-0065, Japan}
}

\date{\normalsize August, 2010}
\maketitle
\thispagestyle{empty}

\begin{abstract}
\normalsize
We investigate coordinate independent SO(9) vector states in SU(2) Matrix theory.
There are 36 vector states, and we determine what representations of SU(2) they are
decomposed into. Among them we find a unique set of states transforming in adjoint representation.
We show that this set of states can appear as the linear term in the coordinate matrices 
in Taylor expansion of zero energy bound state wavefunction around the origin i.e. 
it satisfies the condition of full supersymmetry.
\end{abstract}
\end{titlepage}

\clearpage
%%%%%%%%%%%%%%%%%%%%%%%%%%%%%%%%%%%%%%%%%%%%%%%%%%%%%%%%%%%%%%%%%%%
\section{Introduction}

Construction of zero energy normalizable wavefunction is one of the long standing problems in Matrix theory,
and attempts from various viewpoint have been made. One of such attempts is to investigate 
Taylor expansion in the coordinate matrices $X_i^a$ ($i=1,\dots 9$) around the origin.
The first term of the expansion, which is independent of $X_i^a$,
must be gauge and SO(9) invariant\cite{hh02}. Construction of such states is not an easy mathematical problem,
and in the case of SU(2) gauge group, it has been constructed in \cite{wosiek05,hlt08},
and shown that it is unique\cite{hlt08,ht10}. 

To construct higher terms in the expansion we have to classify more general states independent of $X_i^a$,
This is important not only for constructing zero energy bound states,
but also for constructing general gauge invariant wavefunctions in Matrix theory. 
The number of states independent of $X_i^a$ is counted as follows: 
the fermionic matrix $\theta_\alpha^a$ has $16\times(N^2-1)$
components for SU($N$) gauge group, and out of them we can construct $8(N^2-1)$ creation operators.
Therefore we have $2^{8(N^2-1)}$ states. To deal with this large number of states,
we need a systematic way of classification.
This is also an interesting mathematical problem in its own right.

As a modest step toward this, in this paper we consider SO(9) vector states
independent of $X_i^a$ in SU(2) Matrix theory.
Such states appear in Taylor expansion of the zero energy wavefunction $\ket{\psi}$ as follows:
As is always in supersymmetric systems,
$\ket{\psi}$ is necessarily annihilated by supercharge $Q$, which is given by
\beq
Q={\rm tr}\left[\Pi^i\gamma^i\theta+\frac{i}{2}[X^i,X^j]\gamma^{ij}\theta\right],
\eeq
where $\Pi^i$ are the momenta conjugate to $X^i$.
$\ket{\psi}$ must be gauge invariant, and moreover
it has been shown in \cite{hh02} that $\ket{\psi}$ is SO(9) invariant.
Therefore its Taylor expansion in $X_i^a$ is given by
\bea
\ket{\psi} & = & \ket{\phi}+X_i^a\ket{\phi_i^a}+X_{i_1}^{a_1}X_{i_2}^{a_2}\ket{\phi_{i_1i_2}^{a_1a_2}}+\dots
\nn
& = & \sum_{n=0}^\infty X_{i_1}^{a_1}\dots X_{i_n}^{a_n}\ket{\phi_{i_1\dots i_n}^{a_1\dots a_n}},
\label{taylorexp}
\eea
where $\ket{\phi_{i_1\dots i_n}^{a_1\dots a_n}}$ are states constructed by acting creation operators
made of $\theta_\alpha^a$ on the vacuum for those operators.
The leading term $\ket{\phi}$ has been constructed in \cite{wosiek05,hlt08}.

The equation $Q\ket{\psi}=0$ is decomposed into three independent sequences $m=0,1,2$ relating
$\ket{\phi_{i_1\dots i_{3n+m}}^{a_1\dots a_{3n+m}}}$ and 
$\ket{\phi_{i_1\dots i_{3(n+1)+m}}^{a_1\dots a_{3(n+1)+m}}}$,
and first two of those equations contain only one $\ket{\phi_{i_1\dots i_n}^{a_1\dots a_n}}$:
\bea
& & \gamma^i\theta^a\ket{\phi_i^a}=0,
\label{firstterm} \\
& & \gamma^{i_1}\theta^{a_1}\ket{\phi_{i_1i_2}^{a_1a_2}}=0, \\
& & {}[X^i,X^j]^a X_{i_1}^{a_1}\dots X_{i_{3n+m}}^{a_{3n+m}}\gamma^{ij}\theta^a\ket{\phi_{i_1\dots i_{3n+m}}^{a_1\dots a_{3n+m}}} \nn
& & =2(3(n+1)+m)X_{i_2}^{a_2}\dots X_{i_{3(n+1)+m}}^{a_{3(n+1)+m}}\gamma^{i_1}\theta^{a_1}
\ket{\phi_{i_1\dots i_{3(n+1)+m}}^{a_1\dots a_{3(n+1)+m}}},
\eea
where $n=0,1,2,\dots$
The first equation \eqref{firstterm} restricts the first order term in $X_i^a$.
Since $\ket{\phi_i^a}$ is an SO(9) vector, in this paper we investigate gauge transformation
property of SO(9) vector states. In section 3 we enumerate SO(9) vector states and
show that these states are classified into 15-, 11-, 7-, and 3-dimensional representation of SU(2).
The 3-dimensional i.e. adjoint representation is a unique candidate of $\ket{\phi_i^a}$.
It is shown that it satisfies \eqref{firstterm}, and therefore $\ket{\phi_i^a}$ can be proportional
to $\ket{\phi_i^a}$ with a nonzero coefficient.
Section 2 is for preparing notation and method for investigating representations of SU(2)
through reviewing analysis of SO(9) singlet case in \cite{ht10}.
Calculations are made with the help of symbolic manipulation program Mathematica and
the package for $\gamma$-matrix algebra GAMMA\cite{gran01}.

%%%%%%%%%%%%%%%%%%%%%%%%%%%%%%%%%%%%%%%%%%%%%%%%%%%%%%%%%%%%%%%%%%%
\section{SO(9) singlets}

Before discussing SO(9) vector states, we quickly review SO(9) singlet states and show how
we can extract representations of the gauge group from them.

States are constructed by acting creation operators made of $\theta^a_\alpha$ on the vacuum.
When we fix the adjoint index $a$ of the gauge group, we obtain states in the following representations of SO(9):
\bea
\text{symmetric traceless repr.} & & \ket{ij}, \nn
\text{antisymmetric repr.} & & \ket{ijk}, \nn
\text{vector-spinor repr.} & & \ket{\alpha i}. \nonumber
\eea
These satisfy $\ket{ii}=0$ and $(\gamma^i)_{\alpha\beta}\ket{\beta i}=0$ (Rarita-Schwinger condition).

Action of $\theta_\alpha$ on these states (intertwining relations\cite{hlt08,ht10}) in our
notation\footnote{Our normalization is related to that in \cite{hlt08,ht10} by
$\ket{ij}_\text{there}=-\frac{3}{2}\ket{ij}_\text{here},
\ket{ijk}_\text{there}=3\sqrt{\frac{3}{2}}i\ket{ijk}_\text{here}$, and
$\ket{i\alpha}_\text{there}=\ket{\alpha i}_\text{here}$.} is
\bea
\theta_\alpha\ket{ij} & = & -\frac{1}{3}
\big[(\gamma^i)_{\alpha\beta}\ket{\beta j}+(\gamma^j)_{\alpha\beta}\ket{\beta i}\big],
\label{itwr1} \\
\theta_\alpha\ket{ijk} & = & \frac{1}{\sqrt{3}}(\gamma^{[ij})_{\alpha\beta}\ket{\beta k]},
\label{itwr2} \\
\theta_\alpha\ket{\beta i} & = & -\frac{3}{4}(\gamma^j)_{\alpha\beta}\ket{ij}
-\frac{\sqrt{3}}{24}(\gamma^{jkl}\gamma^i-9\delta^{ij}\gamma^{kl})_{\alpha\beta}\ket{jkl},
\label{itwr3} 
\eea
where $(\gamma^i)_{\alpha\beta}$ are $16\times 16$ real and symmetric SO(9) gamma matrices,
and $[ijk]$ denotes antisymmetrization with the factor $1/3!$.

Nonzero inner products of these states are 
\bea
\vev{ij|kl} & = & \frac{1}{2}(\delta_{ik}\delta_{jl}+\delta_{jk}\delta_{il})-\frac{1}{9}\delta_{ij}\delta_{kl},
\\
\vev{ijk|lmn} & = & \delta_l^{[i}\delta_m^j\delta_n^{k]},
\\
\vev{\alpha i|\beta j} & = & \delta_{ij}\delta_{\alpha\beta}-\frac{1}{8}(\gamma^{ij})_{\alpha\beta}.
\eea
We have to take the adjoint indices into account and we obtain the above states for each index.
For index $a$ those states are denoted by $\ket{*}_a$. Note that $\ket{*}_a$ does not necessarily
transform as adjoint representation of the gauge group.

Full states are constructed by taking products of states from each index, and in the case of SU(2) gauge group,
SO(9) singlet states are classified into the following 14 states $\ket{I} (I=1,2,\dots,14)$
\footnote{Note that for later convenience we put an additional sign factor on $S_7$ in \cite{ht10}: $\ket{7}=-S_7$.}
\cite{ht10}:
\beq
\ket{1}=\ket{ij}_1\ket{jk}_2\ket{ki}_3,
\eeq
\beq
\ket{2}=\epsilon^{ijklmnpqr}\ket{ijk}_1\ket{lmn}_2\ket{pqr}_3,
\eeq
\bea
\ket{3}=\ket{ikl}_1\ket{jkl}_2\ket{ij}_3,~
\ket{4}=\ket{jkl}_1\ket{ij}_2\ket{ikl}_3,~
\ket{5}=\ket{ij}_1\ket{ikl}_2\ket{jkl}_3,
\eea
\bea
\ket{6}=\ket{\alpha i}_1\ket{\alpha j}_2\ket{ij}_3,~
\ket{7}=-\ket{\alpha j}_1\ket{ij}_2\ket{\alpha i}_3,~
\ket{8}=\ket{ij}_1\ket{\alpha i}_2\ket{\alpha j}_3,
\eea
\bea
\ket{9}=(\gamma^k)_{\alpha\beta}\ket{\alpha i}_1\ket{\beta j}_2\ket{ijk}_3,\quad
\ket{10}=(\gamma^k)_{\alpha\beta}\ket{\alpha i}_1\ket{ijk}_2\ket{\beta j}_3,\nn
\ket{11}=(\gamma^k)_{\alpha\beta}\ket{ijk}_1\ket{\alpha i}_2\ket{\beta j}_3,
\eea
\bea
\ket{12}=(\gamma^{jkl})_{\alpha\beta}\ket{\alpha i}_1\ket{\beta i}_2\ket{jkl}_3,\quad
\ket{13}=(\gamma^{jkl})_{\alpha\beta}\ket{\alpha i}_1\ket{jkl}_2\ket{\beta i}_3,\nn
\ket{14}=(\gamma^{jkl})_{\alpha\beta}\ket{jkl}_1\ket{\alpha i}_2\ket{\beta i}_3.
\eea
At first glance it is not clear how these states transform under gauge transformation.
It is read off by acting generators of the gauge group
$G^a=\frac{i}{2}f_{abc}\theta^b_\alpha\theta^c_\alpha$.
Since $G^a$ is always proportional to the structure constant $f_{123}$ in the SU(2) case,
we use $g^a=G^a/if_{123}$ instead of $G^a$ in the following.
Representation matrix $M_{IJ}$ of $g^1$ on these states: 
\beq
g^1\ket{I}=M_{IJ}\ket{J},
\eeq
can be computed using the intertwining relations \eqref{itwr1}-\eqref{itwr3},
and its components are real in our notation.
Explicit expression of $M_{IJ}$ may be read off from the result given in \cite{ht10}.
In our notation $M$ is given by
\footnote{There are some discrepancies between our result \eqref{repm0} and that read off from \cite{ht10}.
We think our result is correct because it gives an appropriate eigenvalue spectrum \eqref{es0}
and passes the test of hermiticity. However both results give the same eigenvectors of eigenvalue 0
and the conclusion about SO(9)$\times$SU(2) singlet is common.
We thank M.\ Hynek and M.\ Trzetrzelewski for 
correspondence about this point.}
\beq
{\tiny
M=\left(\begin{array}{rrrrrrrrrrrrrr}
0 & 0 &  0 & 0 & 0 &  0 & 0 & \frac{13}{9} &  0 & 0 & 0 &  0 & 0 & 0 \\
0 & 0 &  0 & 0 & 0 &  0 & 0 & 0 &  0 & 0 & -48 &  0 & 0 & 8 \\
0 & 0 &  0 & 0 & 0 &  0 & 0 & 0 &  0 & 0 & -\frac{2}{\sqrt{3}} &  0 & 0 & \frac{1}{9\sqrt{3}} \\
0 & 0 &  0 & 0 & 0 &  0 & 0 & 0 &  0 & 0 & \frac{2}{\sqrt{3}} &  0 & 0 & -\frac{1}{9\sqrt{3}} \\
0 & 0 &  0 & 0 & 0 &  0 & 0 & \frac{2}{3} &  0 & 0 & 0 &  0 & 0 & 0 \\
0 & 0 &  0 & 0 & 0 &  0 & \frac{1}{2} & 0 &  0 & -\frac{1}{2\sqrt{3}} & 0 &  0 & -\frac{7}{12\sqrt{3}} & 0 \\
0 & 0 &  0 & 0 & 0 &  -\frac{1}{2} & 0 & 0 &  -\frac{1}{2\sqrt{3}} & 0 & 0 &  -\frac{7}{12\sqrt{3}} & 0 & 0 \\
-9 & 0 &  0 & 0 & -\frac{9}{2} &  0 & 0 & 0 &  0 & 0 & 0 &  0 & 0 & 0 \\
0 & 0 &  0 & 0 & 0 &  0 & \sqrt{3} & 0 &  0 & -2 & 0 &  0 & -\frac{1}{6} & 0 \\
0 & 0 &  0 & 0 & 0 &  \sqrt{3} & 0 & 0 &  2 & 0 & 0 &  \frac{1}{6} & 0 & 0 \\
0 & \frac{1}{12} &  6\sqrt{3} & -6\sqrt{3} & 0 &  0 & 0 & 0 &  0 & 0 & 0 &  0 & 0 & 0 \\
0 & 0 &  0 & 0 & 0 &  0 & 21\sqrt{3} & 0 &  0 & -15 & 0 &  0 & -\frac{1}{2} & 0 \\
0 & 0 &  0 & 0 & 0 &  21\sqrt{3} & 0 & 0 &  15 & 0 & 0 &  \frac{1}{2} & 0 & 0 \\
0 & -\frac{5}{4} &  -9\sqrt{3} & 9\sqrt{3} & 0 &  0 & 0 & 0 &  0 & 0 & 0 &  0 & 0 & 0
\end{array}\right)},
\label{repm0}
\eeq
and representation matrices $N_{IJ}$ and $L_{IJ}$ corresponding to $g^2$ and $g^3$ respectively
are given by the following interchange of the indices:
\bea
N_{IJ} & = & M_{IJ}|_{(3,4,5)\rightarrow (4,5,3),(6,7,8)\rightarrow (7,8,6),(9,10,11)\rightarrow (10,11,9),
(12,13,14)\rightarrow (13,14,12)},\\
L_{IJ} & = & M_{IJ}|_{(3,4,5)\rightarrow (5,3,4),(6,7,8)\rightarrow (8,6,7),(9,10,11)\rightarrow (11,9,10),
(12,13,14)\rightarrow (14,12,13)}.
\eea
For example, $N_{3,11}=M_{4,9},L _{1,7}=M_{1,6}$.

For an eigenvector $v_I$ and corresponding eigenvalue $\lambda$ of $M^T$:
\beq
v_IM_{IJ}=\lambda v_J,
\eeq
$v_I\ket{I}$ gives an eigenvector of $g^1$ corresponding to the eigenvalue $\lambda$.
As is well-known, for any unitary representation of SU(2), spectrum of eigenvalues of $G^1/f_{123}=ig^1$
are given as a disjoint union of sets in the form of $(-j,-j+1,\dots,j-1,j)$, where $j$ is 
nonnegative half integer.
Since the spectrum of eigenvalues of $iM_{IJ}$ is given by
\beq
0,0,1,-1,2,-2,3,-3,4,-4,5,-5,6,-6,
\label{es0}
\eeq
there are one singlet and one "spin 6" representation.

One may wonder why $if_{123}M$ is not hermitian, while $G^1$ is a hermitian operator.
This is because the basis $\{\ket{I}; I=1,\dots 14\}$ is not orthonormal.
Indeed, $P_{IJ}\equiv\vev{I|J}$ is given by
\beq
{\tiny
P=\left(\begin{array}{rrrrrrrrrrrrrr}
\frac{1001}{9} & 0 &  0 & 0 & 0 &  0 & 0 & 0 &  0 & 0 & 0 &  0 & 0 & 0 \\
0 & 362880 &  0 & 0 & 0 &  0 & 0 & 0 &  0 & 0 & 0 &  0 & 0 & 0 \\
0 & 0 &  \frac{308}{3} & 0 & 0 &  0 & 0 & 0 &  0 & 0 & 0 &  0 & 0 & 0 \\
0 & 0 &  0 & \frac{308}{3} & 0 &  0 & 0 & 0 &  0 & 0 & 0 &  0 & 0 & 0 \\
0 & 0 &  0 & 0 & \frac{308}{3} &  0 & 0 & 0 &  0 & 0 & 0 &  0 & 0 & 0 \\
0 & 0 &  0 & 0 & 0 &  693 & 0 & 0 &  0 & 0 & 0 &  0 & 0 & 0 \\
0 & 0 &  0 & 0 & 0 &  0 & 693 & 0 &  0 & 0 & 0 &  0 & 0 & 0 \\
0 & 0 &  0 & 0 & 0 &  0 & 0 & 693 &  0 & 0 & 0 &  0 & 0 & 0 \\
0 & 0 &  0 & 0 & 0 &  0 & 0 & 0 &  1071 & 0 & 0 &  2646 & 0 & 0 \\
0 & 0 &  0 & 0 & 0 &  0 & 0 & 0 &  0 & 1071 & 0 &  0 & 2646 & 0 \\
0 & 0 &  0 & 0 & 0 &  0 & 0 & 0 &  0 & 0 & 1071 &  0 & 0 & 2646 \\
0 & 0 &  0 & 0 & 0 &  0 & 0 & 0 &  2646 & 0 & 0 &  72576 & 0 & 0 \\
0 & 0 &  0 & 0 & 0 &  0 & 0 & 0 &  0 & 2646 & 0 &  0 & 72576 & 0 \\
0 & 0 &  0 & 0 & 0 &  0 & 0 & 0 &  0 & 0 & 2646 &  0 & 0 & 72576
\end{array}\right)},
\eeq
and we have checked that $if_{123}M_{JK}P_{IK}=\vev{I|G^1|J}$ is hermitian.

$g^1$ has two eigenvectors corresponding to the eigenvalue 0:
\bea
v^{(1)} & = & (0, 0, 1, 1, 0, 0, 0, 0, 0, 0, 0, 0, 0, 0), \\
v^{(2)} & = & (-6/13,0,0,0,1, 0, 0, 0, 0, 0, 0, 0, 0, 0),
\eea
and only the combination $v^{(1)}+v^{(2)}$ gives a state of eigenvalue 0 for $g^2$ and $g^3$.
In other words, the following state $\ket{S}$:
\beq
\ket{S}=-\frac{6}{13}\ket{ij}_1\ket{jk}_2\ket{ki}_3
+ \ket{ikl}_1\ket{jkl}_2\ket{ij}_3+\ket{ikl}_1\ket{ij}_2\ket{jkl}_3+\ket{ij}_1\ket{ikl}_2\ket{jkl}_3
\label{singlet}
\eeq
is the unique SU(2)$\times$SO(9) singlet state\cite{hlt08,ht10}, and $\ket{\phi}$ must be proportional to $\ket{S}$.

To construct the spin 6 representation, and for later use, we consider general representations of integer spins.
Such representations can be constructed as symmetric tensor products of adjoint (spin 1) representations
with the tracelessness condition. i.e.
a $(2n+1)$-dimensional (spin $n$) representation is given by states
$\ket{a_1a_2\dots a_n}$ which satisfy
\beq
\ket{a_1a_2\dots a_{m-1}a_ma_{m+1}\dots a_n}=\ket{a_ma_2\dots a_{m-1}a_1a_{m+1}\dots a_n},\quad
\ket{aaa_3\dots a_n}=0,
\eeq
and we define indices $\pm$ as follows:
\beq
\ket{\pm a_2\dots a_n}=\frac{1}{\sqrt{2}}(\ket{2a_2\dots a_n}\pm i\ket{3a_2\dots a_n}).
\eeq
Then, from the tracelessness condition, $\ket{+-a_3\dots a_n}=-\frac{1}{2}\ket{11a_3\dots a_n}$.
This representation has the unique eigenvectors of $ig^1$ corresponding to the eigenvalue $0, m$, and $-m$:
\begin{align}
ig^1\ket{1\dots 1} & = 0,\\
ig^1|1\dots 1\underbrace{+\dots +}_m\rangle & = -m|1\dots 1\underbrace{+\dots +}_m\rangle,\\
ig^1|1\dots 1\underbrace{-\dots -}_m\rangle & = m|1\dots 1\underbrace{-\dots -}_m\rangle.
%ig^1(\ket{11\dots 12}\pm i\ket{11\dots 13})=\mp(\ket{11\dots 12}\pm i\ket{11\dots 13}).
\end{align}
Operators $ig^\pm\equiv\frac{i}{\sqrt{2}}(g^2\pm ig^3)$ increase and decrease numbers of $+$ and $-$ in these states:
\begin{align}
ig^+|+\dots +\underbrace{1\dots 1}_m\rangle & = m|+\dots +\underbrace{1\dots 1}_{m-1}\rangle,\\
ig^+|1\dots 1\underbrace{-\dots -}_m\rangle & = (-1)^mm|1\dots 1\underbrace{-\dots -}_{m-1}\rangle,\\
ig^-|-\dots -\underbrace{1\dots 1}_m\rangle & = (-1)^mm|-\dots -\underbrace{1\dots 1}_{m-1}\rangle,\\
ig^-|1\dots 1\underbrace{+\dots +}_m\rangle & = m|1\dots 1\underbrace{+\dots +}_{m-1}\rangle.
%ig^1(\ket{11\dots 12}\pm i\ket{11\dots 13})=\mp(\ket{11\dots 12}\pm i\ket{11\dots 13}).
\end{align}
Now the construction of the spin 6 representation is straightforward.
The eigenvector of $iM^T$ corresponding to the eigenvalue 6 is
\beq 
(0, -3i, -108\sqrt{3}i, 108\sqrt{3}i, 0, 0, 0, 0, 0, 0, -96, 0, 0,8),
\eeq
which corresponds to $\ket{------}$. The rest of states are constructed by acting $g^+$ successively:
\begin{align}
\ket{------}= & -3i\ket{2}-108\sqrt{3}i\ket{3}+108\sqrt{3}i\ket{4}-96\ket{11}+8\ket{14},\notag\\
\ket{-----1}= & \frac{1}{6}ig^+\ket{------}, \notag\\
\ket{----11}= & -\frac{1}{5}ig^+\ket{-----1}, \notag\\
\ket{---111}= & \frac{1}{4}ig^+\ket{----11}, \notag\\
\vdots &
\end{align}
Since it is not illuminating, we do not write down explicit expressions of all of these states.
We only give $\ket{111111}$, which is in the kernel of $ig^1$:
\bea 
\ket{111111} & = & -\frac{1}{6!}(ig^+)^6\ket{------} \nn
& = & 324\sqrt{3}i\ket{1}+432\sqrt{3}i\ket{3}+432\sqrt{3}i\ket{4}-702\sqrt{3}i\ket{5} \nn
& = & 432\sqrt{3}i\left(-\frac{13}{8}v^{(1)}_I\ket{I}+v^{(2)}_I\ket{I}\right),
\eea
and is orthogonal to the state \eqref{singlet}.
These states may appear in Taylor expansion of zero energy wavefunction in the form of
$(X_{i_1}^{a_1}X_{i_1}^{a_2})(X_{i_2}^{a_3}X_{i_2}^{a_4})(X_{i_3}^{a_5}X_{i_3}^{a_6})\ket{a_1a_2a_3a_4a_5a_6}$.

%%%%%%%%%%%%%%%%%%%%%%%%%%%%%%%%%%%%%%%%%%%%%%%%%%%%%%%%%%%%%%%%%%%
\section{SO(9) vectors}

In this section we investigate SO(9) vector states, and as in the previous section we analyze their
transformation property under gauge transformation from the viewpoint of the eigenvalue spectrum of the 
representation matrix.

There are 36 sets of states in SO(9) vector representation, and they are denoted by $\ket{I,a,i}$, where
$I=1,2,\dots,12$. When $\ket{I,1,i}$ takes the form of $\ket{*_1}_1\ket{*_2}_2\ket{*_3}_3$,
$\ket{I,2,i}$ and $\ket{I,3,i}$ are given by $\ket{*_1}_2\ket{*_2}_3\ket{*_3}_1$ and 
$\ket{*_1}_3\ket{*_2}_1\ket{*_3}_2$ respectively.
Note that $\ket{I,a,i}$ does not necessarily transform as SU(2) adjoint representation.

Explicitly, $\ket{I,a,i}$ are defined as follows:
\beq
\ket{1,1,i}=\ket{ijk}_1\ket{jl}_2\ket{kl}_3,~
\ket{1,2,i}=\ket{kl}_1\ket{ijk}_2\ket{jl}_3,~
\ket{1,3,i}=\ket{jl}_1\ket{kl}_2\ket{ijk}_3,
\label{ket1ai}
\eeq
\bea
\ket{2,1,i}=\ket{ijk}_1\ket{jlm}_2\ket{klm}_3,\quad
\ket{2,2,i}=\ket{klm}_1\ket{ijk}_2\ket{jlm}_3,\nn
\ket{2,3,i}=\ket{jlm}_1\ket{klm}_2\ket{ijk}_3,
\eea
\bea
\ket{3,1,i}=\ket{ij}_1\ket{\alpha k}_2\ket{\beta k}_3(\gamma^j)_{\alpha\beta},\quad
\ket{3,2,i}=-\ket{\beta k}_1\ket{ij}_2\ket{\alpha k}_3(\gamma^j)_{\alpha\beta},\nn
\ket{3,3,i}=\ket{\alpha k}_1\ket{\beta k}_2\ket{ij}_3(\gamma^j)_{\alpha\beta},
\eea
\bea
\ket{4,1,i}=\ket{jk}_1\ket{\alpha i}_2\ket{\beta j}_3(\gamma^k)_{\alpha\beta},\quad
\ket{4,2,i}=-\ket{\beta j}_1\ket{jk}_2\ket{\alpha i}_3(\gamma^k)_{\alpha\beta},\nn
\ket{4,3,i}=\ket{\alpha i}_1\ket{\beta j}_2\ket{jk}_3(\gamma^k)_{\alpha\beta},
\eea
\bea
\ket{5,1,i}=\ket{jk}_1\ket{\alpha j}_2\ket{\beta i}_3(\gamma^k)_{\alpha\beta},\quad
\ket{5,2,i}=-\ket{\beta i}_1\ket{jk}_2\ket{\alpha j}_3(\gamma^k)_{\alpha\beta},\nn
\ket{5,3,i}=\ket{\alpha j}_1\ket{\beta i}_2\ket{jk}_3(\gamma^k)_{\alpha\beta},
\eea
\bea
\ket{6,1,i}=\ket{jk}_1\ket{\alpha j}_2\ket{\beta k}_3(\gamma^i)_{\alpha\beta},\quad
\ket{6,2,i}=-\ket{\beta k}_1\ket{jk}_2\ket{\alpha j}_3(\gamma^i)_{\alpha\beta},\nn
\ket{6,3,i}=\ket{\alpha j}_1\ket{\beta k}_2\ket{jk}_3(\gamma^i)_{\alpha\beta},
\eea
\bea
\ket{7,1,i}=\ket{ijk}_1\ket{\alpha j}_2\ket{\alpha k}_3,\quad
\ket{7,2,i}=-\ket{\alpha k}_1\ket{ijk}_2\ket{\alpha j}_3,\nn
\ket{7,3,i}=\ket{\alpha j}_1\ket{\alpha k}_2\ket{ijk}_3,
\eea
\bea
\ket{8,1,i}=\ket{ijk}_1\ket{\alpha l}_2\ket{\beta l}_3(\gamma^{jk})_{\alpha\beta},\quad
\ket{8,2,i}=-\ket{\beta l}_1\ket{ijk}_2\ket{\alpha l}_3(\gamma^{jk})_{\alpha\beta},\nn
\ket{8,3,i}=\ket{\alpha l}_1\ket{\beta l}_2\ket{ijk}_3(\gamma^{jk})_{\alpha\beta},
\eea
\bea
\ket{9,1,i}=\ket{jkl}_1\ket{\alpha i}_2\ket{\beta l}_3(\gamma^{jk})_{\alpha\beta},\quad
\ket{9,2,i}=-\ket{\beta l}_1\ket{jkl}_2\ket{\alpha i}_3(\gamma^{jk})_{\alpha\beta},\nn
\ket{9,3,i}=\ket{\alpha i}_1\ket{\beta l}_2\ket{jkl}_3(\gamma^{jk})_{\alpha\beta},
\eea
\bea
\ket{10,1,i}=\ket{jkl}_1\ket{\alpha l}_2\ket{\beta i}_3(\gamma^{jk})_{\alpha\beta},\quad
\ket{10,2,i}=-\ket{\beta i}_1\ket{jkl}_2\ket{\alpha l}_3(\gamma^{jk})_{\alpha\beta},\nn
\ket{10,3,i}=\ket{\alpha l}_1\ket{\beta i}_2\ket{jkl}_3(\gamma^{jk})_{\alpha\beta},
\eea
\bea
\ket{11,1,i}=\ket{jkl}_1\ket{\alpha j}_2\ket{\beta k}_3(\gamma^{il})_{\alpha\beta},\quad
\ket{11,2,i}=-\ket{\beta k}_1\ket{jkl}_2\ket{\alpha j}_3(\gamma^{il})_{\alpha\beta},\nn
\ket{11,3,i}=\ket{\alpha j}_1\ket{\beta k}_2\ket{jkl}_3(\gamma^{il})_{\alpha\beta},
\eea
\bea
\ket{12,1,i}=\ket{jkl}_1\ket{\alpha m}_2\ket{\beta m}_3(\gamma^{ijkl})_{\alpha\beta},\quad
\ket{12,2,i}=-\ket{\beta m}_1\ket{jkl}_2\ket{\alpha m}_3(\gamma^{ijkl})_{\alpha\beta},\nn
\ket{12,3,i}=\ket{\alpha m}_1\ket{\beta m}_2\ket{jkl}_3(\gamma^{ijkl})_{\alpha\beta}.
\label{ket12ai}
\eea
In the following, components of vectors corresponding to the above states are placed
in the order in \eqref{ket1ai}-\eqref{ket12ai}. For example, 
$(v_{(1,1)},v_{(1,2)},v_{(1,3)},v_{(2,1)},v_{(2,2)},v_{(2,3)},\dots)$ for $v_{(Ia)}$.
Under this rule, representation matrix of $g^1$:
\beq
g^1\ket{I,a,i}=M_{(Ia),(Jb)}\ket{J,b,i},
\eeq
which is calculated by using the intertwining relations, is given as follows:
\bea
M_{(1,1),{(Jb)}} & = & (0,0,0, 0,0,0, 0,0,0, 0,0,0, 0,0,0, 0,0,0, \nn
& & 13/9,0,0, 1/9,0,0, 0,0,0, 0,0,0, 0,0,0, 0,0,0), \nn
M_{(1,2),{(Jb)}} & = & (0,0,0, 0,0,0, 1/(9\sqrt{3}),0,0, -7/(9\sqrt{3}),0,0, -1/(9\sqrt{3}),0,0, 8/(9\sqrt{3}),0,0, \nn
& & 0,0,0, 0,0,0, 0,0,0, 0,0,0, 0,0,0, 0,0,0), \nn
M_{(1,3),{(Jb)}} & = & (0,0,0, 0,0,0, -1/(9\sqrt{3}),0,0, 1/(9\sqrt{3}),0,0, 7/(9\sqrt{3}),0,0, -8/(9\sqrt{3}),0,0, \nn
& & 0,0,0, 0,0,0, 0,0,0, 0,0,0, 0,0,0, 0,0,0), \nn
M_{(2,1),{(Jb)}} & = & (0,0,0, 0,0,0, 0,0,0, 0,0,0, 0,0,0, 0,0,0, \nn
& & 22/27,0,0, 10/27,0,0, 0,0,0, 0,0,0, 0,0,0, 0,0,0), \nn
M_{(2,2),{(Jb)}} & = & (0,0,0, 0,0,0, 0,0,0, 0,0,0, 0,0,0, 0,0,0, \nn
& & 8/27,0,0, 2/27,0,0, 5/27,0,0, 1/27,0,0, 4/27,0,0, 1/27,0,0), \nn
M_{(2,3),{(Jb)}} & = & (0,0,0, 0,0,0, 0,0,0, 0,0,0, 0,0,0, 0,0,0, \nn
& & 8/27,0,0, 2/27,0,0, 1/27,0,0, 5/27,0,0, -4/27,0,0, -1/27,0,0), \nn
M_{(3,1),{(Jb)}} & = & (0,0,0, 0,0,0, 0,0,0, 0,0,0, 0,0,0, 0,0,0, \nn
& & 0,0,0, 0,0,0, 0,0,0, 0,0,0, 0,0,0, 0,0,0), \nn
M_{(3,2),{(Jb)}} & = & (0,0,0, 0,0,0, 0,0,0, 0,0,0, 0,0,9/4, 0,0,-1/2, \nn
& & 0,0,-\sqrt{3}/2, 0,0,\sqrt{3}/12, 0,0,0, 0,0,-17\sqrt{3}/24, 0,0,\sqrt{3}/2, 0,0,-\sqrt{3}/36), \nn
M_{(3,3),{(Jb)}} & = & (0,0,0, 0,0,0, 0,0,0, 0,-9/4,0, 0,0,0, 0,1/2,0, \nn
& & 0,-\sqrt{3}/2,0, 0,\sqrt{3}/12,0, 0,-17\sqrt{3}/24,0, 0,0,0, 0,-\sqrt{3}/2,0, 0,\sqrt{3}/36,0), \nn
M_{(4,1),{(Jb)}} & = & (0,6\sqrt{3},0, 0,0,0, 0,0,0, 0,0,0, 0,0,0, \nn
& & 0,0,0, 0,0,0, 0,0,0, 0,0,0, 0,0,0, 0,0,0, 0,0,0), \nn
M_{(4,2),{(Jb)}} & = & (0,0,0, 0,0,0, 0,0,9/4, 0,0,0, 0,0,0, 0,0,0, \nn
& & 0,0,2\sqrt{3}/3, 0,0,-7\sqrt{3}/12, 0,0,-\sqrt{3}/6, 0,0,\sqrt{3}/6, 0,0,-1/\sqrt{3}, 0,0,\sqrt{3}/24), \nn
M_{(4,3),{(Jb)}} & = & (0,0,0, 0,0,0, 0,0,0, 0,0,0, 0,-7/4,0, 0,0,0, \nn
& & 0,0,0, 0,0,0, 0,0,0, 0,-11\sqrt{3}/8,0, 0,0,0, 0,0,0), \nn
M_{(5,1),{(Jb)}} & = & (0,0,-6\sqrt{3}, 0,0,0, 0,0,0, 0,0,0, 0,0,0, 0,0,0, \nn
& & 0,0,0, 0,0,0, 0,0,0, 0,0,0, 0,0,0, 0,0,0), \nn
M_{(5,2),{(Jb)}} & = & (0,0,0, 0,0,0, 0,0,0, 0,0,7/4, 0,0,0, 0,0,0, 0,0,0, \nn
& & 0,0,0, 0,0,-11\sqrt{3}/8, 0,0,0, 0,0,0, 0,0,0), \nn
M_{(5,3),{(Jb)}} & = & (0,0,0, 0,0,0, 0,-9/4,0, 0,0,0, 0,0,0, 0,0,0, \nn
& & 0,2\sqrt{3}/3,0, 0,-7\sqrt{3}/12,0, 0,\sqrt{3}/6,0, 0,-\sqrt{3}/6,0, 0,1/\sqrt{3},0, 0,-\sqrt{3}/24,0), \nn
M_{(6,1),{(Jb)}} & = & (0,-6\sqrt{3},6\sqrt{3}, 0,0,0, 0,0,0, 0,0,0, 0,0,0, 0,0,0, \nn
& & 0,0,0, 0,0,0, 0,0,0, 0,0,0, 0,0,0, 0,0,0), \nn
M_{(6,2),{(Jb)}} & = & (0,0,0, 0,0,0, 0,0,0, 0,0,1/2, 0,0,0, 0,0,-1/2, \nn
& & 0,0,-\sqrt{3}/6, 0,0,-7\sqrt{3}/12, 0,0,-\sqrt{3}/12, 0,0,0, 0,0,-\sqrt{3}/6, 0,0,-7\sqrt{3}/36), \nn
M_{(6,3),{(Jb)}} & = & (0,0,0, 0,0,0, 0,0,0, 0,0,0, 0,-1/2,0, 0,1/2,0, \nn
& & 0,-\sqrt{3}/6,0, 0,-7\sqrt{3}/12,0, 0,0,0, 0,-\sqrt{3}/12,0, 0,\sqrt{3}/6,0, 0,7\sqrt{3}/36,0), \nn
M_{(7,1),{(Jb)}} & = & (-9,0,0, -15/2,0,0, 0,0,0, 0,0,0, 0,0,0, 0,0,0, \nn
& & 0,0,0, 0,0,0, 0,0,0, 0,0,0, 0,0,0, 0,0,0), \nn
M_{(7,2),{(Jb)}} & = & (0,0,0, 0,0,0, 0,0,1/(4\sqrt{3}), 0,0,-1/(4\sqrt{3}), 0,0,-1/(4\sqrt{3}), 0,0,1/(2\sqrt{3}), \nn
& & 0,0,1/6, 0,0,-1/3, 0,0,-1/24, 0,0,17/24, 0,0,-1/6, 0,0,1/8), \nn
M_{(7,3),{(Jb)}} & = & (0,0,0, 0,0,0, 0,1/(4\sqrt{3}),0, 0,-1/(4\sqrt{3}),0, 0,-1/(4\sqrt{3}),0, 0,1/(2\sqrt{3}),0, \nn
& & 0,-1/6,0, 0,1/3,0, 0,-17/24,0, 0,1/24,0, 0,-1/6,0, 0,1/8,0), \nn
M_{(8,1),{(Jb)}} & = & (-18,0,0, -51,0,0, 0,0,0, 0,0,0, 0,0,0, 0,0,0, \nn
& & 0,0,0, 0,0,0, 0,0,0, 0,0,0, 0,0,0, 0,0,0), \nn
M_{(8,2),{(Jb)}} & = & (0,0,0, 0,0,0, 0,0,0, 0,0,0, 0,0,3\sqrt{3}/2, 0,0,2\sqrt{3}, \nn
& & 0,0,-14/3, 0,0,1/3, 0,0,-1/3, 0,0,-1/12, 0,0,8/3, 0,0,0), \nn
M_{(8,3),{(Jb)}} & = & (0,0,0, 0,0,0, 0,0,0, 0,3\sqrt{3}/2,0, 0,0,0, 0,2\sqrt{3},0, \nn
& & 0,14/3,0, 0,-1/3,0, 0,1/12,0, 0,1/3,0, 0,8/3,0, 0,0,0), \nn
M_{(9,1),{(Jb)}} & = & (0,0,0, 0,-18,0, 0,0,0, 0,0,0, 0,0,0, 0,0,0, \nn
& & 0,0,0, 0,0,0, 0,0,0, 0,0,0, 0,0,0, 0,0,0), \nn
M_{(9,2),{(Jb)}} & = & (0,0,0, 0,0,0, 0,0,3\sqrt{3}/2, 0,0,0, 0,0,0, \nn
& & 0,0,0, 0,0,20/3, 0,0,1/6, 0,0,-1/3, 0,0,1/3, 0,0,-2/3, 0,0,1/12), \nn
M_{(9,3),{(Jb)}} & = & (0,0,0, 0,0,0, 0,0,0, 0,0,0, 0,7\sqrt{3}/2,0, \nn
& & 0,0,0, 0,0,0, 0,0,0, 0,0,0, 0,17/4,0, 0,0,0, 0,0,0), \nn
M_{(10,1),{(Jb)}} & = & (0,0,0, 0,0,-18, 0,0,0, 0,0,0, 0,0,0, 0,0,0, \nn
& & 0,0,0, 0,0,0, 0,0,0, 0,0,0, 0,0,0, 0,0,0), \nn
M_{(10,2),{(Jb)}} & = & (0,0,0, 0,0,0, 0,0,0, 0,0,7\sqrt{3}/2, 0,0,0, \nn
& & 0,0,0, 0,0,0, 0,0,0, 0,0,-17/4, 0,0,0, 0,0,0, 0,0,0), \nn
M_{(10,3),{(Jb)}} & = & (0,0,0, 0,0,0, 0,3\sqrt{3}/2,0, 0,0,0, 0,0,0, \nn
& & 0,0,0, 0,-20/3,0, 0,-1/6,0, 0,-1/3,0, 0,1/3,0, 0,-2/3,0, 0,1/12,0), \nn
M_{(11,1),{(Jb)}} & = & (0,0,0, 0,-9,9, 0,0,0, 0,0,0, 0,0,0, 0,0,0, \nn
& & 0,0,0, 0,0,0, 0,0,0, 0,0,0, 0,0,0, 0,0,0), \nn
M_{(11,2),{(Jb)}} & = & (0,0,0, 0,0,0, 0,0,1/(4\sqrt{3}), 0,0,11/(4\sqrt{3}), 0,0,-1/(4\sqrt{3}), 0,0,-10/(4\sqrt{3}), \nn
& & 0,0,-11/6, 0,0,-5/6, 0,0,-29/24, 0,0,17/24, 0,0,-13/6, 0,0,-1/24), \nn
M_{(11,3),{(Jb)}} & = & (0,0,0, 0,0,0, 0,-1/(4\sqrt{3}),0, 0,1/(4\sqrt{3}),0, 0,-11/(4\sqrt{3}),0, 0,10/(4\sqrt{3}),0, \nn
& & 0,-11/6,0, 0,-5/6,0, 0,17/24,0, 0,-29/24,0, 0,13/6,0, 0,1/24,0), \nn
M_{(12,1),{(Jb)}} & = & (0,0,0, 0,-144,144, 0,0,0, 0,0,0, 0,0,0, 0,0,0, \nn
& & 0,0,0, 0,0,0, 0,0,0, 0,0,0, 0,0,0, 0,0,0), \nn
M_{(12,2),{(Jb)}} & = & (0,0,0, 0,0,0, 0,0,0, 0,0,0, 0,0,9\sqrt{3}/2, 0,0,-15\sqrt{3}, \nn
& & 0,0,-29, 0,0,-1/2, 0,0,2, 0,0,-1/4, 0,0,-7, 0,0,-1/2), \nn
M_{(12,3),{(Jb)}} & = & (0,0,0, 0,0,0, 0,0,0, 0,-9\sqrt{3}/2,0, 0,0,0, 0,15\sqrt{3},0, \nn
& & 0,-29,0, 0,-1/2,0, 0,-1/4,0, 0,2,0, 0,7,0, 0,1/2,0).
\eea
This matrix also gives actions of $g^2$ and $g^3$ on $\ket{I,a,i}$:
\beq
g^2\ket{I,a,i}=M_{(I,a-1),(J,b-1)}\ket{J,b,i},\quad g^3\ket{I,a,i}=M_{(I,a+1),(J,b+1)}\ket{J,b,i},
\eeq
where indices $a,b,\dots$ are defined modulo 3.

The eigenvalue spectrum of $iM^T$ is
\bea
& & 0,0,0,0,1,1,1,1,-1,-1,-1,-1,2,2,2,-2,-2,-2,\nn
& & 3,3,3,-3,-3,-3,4,4,-4,-4,5,5,-5,-5,6,-6,7,-7,
\eea
and we see there is one spin 7, one spin 5, one spin 3 and one spin 1 representation.

As a check of the correctness of the above $M$, we have confirmed that
$if_{123}M_{(Jb),(Kc)}P_{(Ia),(Kc)}\delta_{ij} = \vev{I,a,i|G^1|J,b,j}$ is hermitian, where
nonzero components of $P_{(Ia),(Jb)}\delta_{ij}\equiv \vev{I,a,i|J,b,j}=P_{(Jb),(Ia)}\delta_{ji}$ are
\bea
P_{(1,1),(1,1)}=P_{(1,2),(1,2)}=P_{(1,3),(1,3)}=77/3, \nn
P_{(2,1),(2,1)}=P_{(2,2),(2,2)}=P_{(2,3),(2,3)}=196/9, \nn
P_{(2,1),(2,2)}=P_{(2,1),(2,3)}=P_{(2,2),(2,3)}=56/9, \nn
P_{(3,1),(3,1)}=P_{(3,2),(3,2)}=P_{(3,3),(3,3)}=6776/9, \nn
P_{(3,1),(4,1)}=P_{(3,2),(4,2)}=P_{(3,3),(4,3)}=77, \nn
P_{(3,1),(5,1)}=P_{(3,2),(5,2)}=P_{(3,3),(5,3)}=77, \nn
P_{(3,1),(6,1)}=P_{(3,2),(6,2)}=P_{(3,3),(6,3)}=-154/9, \nn
P_{(4,1),(4,1)}=P_{(4,2),(4,2)}=P_{(4,3),(4,3)}=792, \nn
P_{(4,1),(5,1)}=P_{(4,2),(5,2)}=P_{(4,3),(5,3)}=99, \nn
P_{(4,1),(6,1)}=P_{(4,2),(6,2)}=P_{(4,3),(6,3)}=176, \nn
P_{(5,1),(5,1)}=P_{(5,2),(5,2)}=P_{(5,3),(5,3)}=792, \nn
P_{(5,1),(6,1)}=P_{(5,2),(6,2)}=P_{(5,3),(6,3)}=176, \nn
P_{(6,1),(6,1)}=P_{(6,2),(6,2)}=P_{(6,3),(6,3)}=6281/9, \nn
P_{(7,1),(7,1)}=P_{(7,2),(7,2)}=P_{(7,3),(7,3)}=455/3, \nn
P_{(7,1),(8,1)}=P_{(7,2),(8,2)}=P_{(7,3),(8,3)}=322/3, \nn
P_{(7,1),(9,1)}=P_{(7,2),(9,2)}=P_{(7,3),(9,3)}=-28, \nn
P_{(7,1),(10,1)}=P_{(7,2),(10,2)}=P_{(7,3),(10,3)}=-28, \nn
P_{(8,1),(8,1)}=P_{(8,2),(8,2)}=P_{(8,3),(8,3)}=8288/3, \nn
P_{(8,1),(9,1)}=P_{(8,2),(9,2)}=P_{(8,3),(9,3)}=364, \nn
P_{(8,1),(10,1)}=P_{(8,2),(10,2)}=P_{(8,3),(10,3)}=364, \nn
P_{(9,1),(9,1)}=P_{(9,2),(9,2)}=P_{(9,3),(9,3)}=2016, \nn
P_{(9,1),(10,1)}=P_{(9,2),(10,2)}=P_{(9,3),(10,3)}=84, \nn
P_{(9,1),(11,1)}=P_{(9,2),(11,2)}=P_{(9,3),(11,3)}=-420, \nn
P_{(9,1),(12,1)}=P_{(9,2),(12,2)}=P_{(9,3),(12,3)}=1596, \nn
P_{(10,1),(10,1)}=P_{(10,2),(10,2)}=P_{(10,3),(10,3)}=2016, \nn
P_{(10,1),(11,1)}=P_{(10,2),(11,2)}=P_{(10,3),(11,3)}=420, \nn
P_{(10,1),(12,1)}=P_{(10,2),(12,2)}=P_{(10,3),(12,3)}=-1596, \nn
P_{(11,1),(11,1)}=P_{(11,2),(11,2)}=P_{(11,3),(11,3)}=966, \nn
P_{(11,1),(12,1)}=P_{(11,2),(12,2)}=P_{(11,3),(12,3)}=1596, \nn
P_{(12,1),(12,1)}=P_{(12,2),(12,2)}=P_{(12,3),(12,3)}=47712.
\eea

Let us construct representations of the gauge group from higher ones.
First, the eigenvector of $iM^T$ corresponding to the eigenvalue 7:
\bea
u^{(7,7)} & = & (0,0,0, 0,0,0, 0,0,0, 0,0,2i, 0,2,0, 0,0,0, \nn
& & 0,0,0, 0,0,0, 0,0,-i\sqrt{3}, 0,\sqrt{3},0, 0,0,0, 0,0,0)
\eea
gives the state $\ket{-------,i}$ in the spin 7 representation:
\beq
\ket{-------,i}=u^{(7,7)}_{(Ia)}\ket{I,a,i}.
\eeq
Here states in this representation are denoted by $\ket{a_1\dots a_7,i}$.
States in the lower representations are also denoted analogously.
Other states in this representation are obtained by successive action of $ig^+$ on $\ket{-------,i}$.
We write down only the following 3 states in eigenspaces of $iM^T$ corresponding to the eigenvalues 5,3, and 1
respectively:
\bea
\ket{-----11,i} & = & -\frac{1}{42}(ig^+)^2\ket{-------,i}
 = u^{(7,5)}_{(Ia)}\ket{I,a,i}, \\
\ket{---1111,i} & = & \frac{1}{840}(ig^+)^4\ket{-------,i}
 = u^{(7,3)}_{(Ia)}\ket{I,a,i}, \\
\ket{-111111,i} & = & -\frac{1}{5040}(ig^+)^6\ket{-------,i}
 = u^{(7,1)}_{(Ia)}\ket{I,a,i},
\eea
where
\bea
u^{(7,5)} & = & \frac{1}{126}\cdot
(0,0,0, 0,0,0, 0,8,-8i, 0,-8,14i, 0,-14,8i, 0,64,-64i,
\nn & &
0,-32\sqrt{3},-32i\sqrt{3}, 0,-8\sqrt{3},-8i\sqrt{3}, 0,-4\sqrt{3},i\sqrt{3}, 
\nn & &
0,\sqrt{3},-4i\sqrt{3}, 0,16\sqrt{3},-16i\sqrt{3}, 0,4\sqrt{3},-4i\sqrt{3})
, \\
u^{(7,3)} & = &\frac{1}{420\sqrt{3}}\cdot
(0,0,0, 0,0,0, 0,8\sqrt{3},8i\sqrt{3}, 0,88\sqrt{3},-26i\sqrt{3},
\nn & &
0,-26\sqrt{3},88i\sqrt{3},0,64\sqrt{3},64i\sqrt{3}, 0,672,-672i, 
\nn & &
0,168,-168i, 0,-156,15i, 0,-15,156i, 0,48,48i, 0,12,12i)
, \\
u^{(7,1)} & = & \frac{1}{168\sqrt{3}}\cdot
(0,0,0, 0,0,0, 0,-16\sqrt{3},16i\sqrt{3}, 0,304\sqrt{3},-34i\sqrt{3},
\nn & &
0,34\sqrt{3},-304i\sqrt{3}, 0,-128\sqrt{3},128i\sqrt{3}, 0,-576,-576i, 
\nn & &
0,-144,-144i, 0,-408,3i, 0,3,-408i, 0,-96,96i, 0,-24,24i).
\eea
Next let us construct the spin 5 representation. The eigenspace of $iM^T$ corresponding to the eigenvalue 5
is 2 dimensional, and in this space the following $u^{(5,5)}$ is orthogonal to $u^{(7,5)}$
i.e. $\bar{u}^{(7,5)}_{(Ia)}P_{(Ia),(Jb)}u^{(5,5)}_{(Jb)}=0$:
\bea
u^{(5,5)} & = & (0,0,0, 0,0,0, 0,88,-88i, 0,-88,-2i, 0,2,88i, 0,2,-2i,
\nn & & 
0,-118\sqrt{3},-118i\sqrt{3}, 0,29\sqrt{3},29i\sqrt{3}, 0,-44\sqrt{3},11i\sqrt{3},
\nn & & 
0,11\sqrt{3},-44i\sqrt{3}, 0,-58\sqrt{3},58i\sqrt{3}, 0,5\sqrt{3},-5i\sqrt{3}).
\eea
This gives $\ket{-----,i}$ in the spin 5 representation:
\beq
\ket{-----,i}=u^{(5,5)}_{(Ia)}\ket{I,a,i}.
\eeq
Other states in this representation are obtained by successive action of $ig^+$ on $\ket{-----,i}$.
We just show the following 2 states in eigenspaces of $iM^T$ corresponding to the eigenvalues 3 and 1
respectively:
\bea
\ket{---11,i} & = & -\frac{1}{20}(ig^+)^2\ket{-----,i}
 = u^{(5,3)}_{(Ia)}\ket{I,a,i}, \\
\ket{-1111,i} & = & \frac{1}{120}(ig^+)^4\ket{-----,i}
 = u^{(5,1)}_{(Ia)}\ket{I,a,i},
\eea
where
\bea
u^{(5,3)} & = & \frac{1}{20}\cdot(0,0,0, 0,0,0, 0,320,320i, 0,400,130i, 0,130,400i, 0,-950,-950i, 
\nn & &
0,354\sqrt{3},-354i\sqrt{3}, 0,-87\sqrt{3},87i\sqrt{3}, 0,0,-99i\sqrt{3}, 0,99\sqrt{3},0,
\nn & &
0,-114\sqrt{3},-114i\sqrt{3}, 0,69\sqrt{3},69i\sqrt{3})
, \\
u^{(5,1)} & = & \frac{1}{10}\cdot(0,0,0, 0,0,0, 0,500,-500i, 0,-140,770i, 0,-770,140i, 0,490,-490i,
\nn & &
0,826\sqrt{3},826i\sqrt{3}, 0,-203\sqrt{3},-203i\sqrt{3}, 0,-66\sqrt{3},297i\sqrt{3},
\nn & &
0,297\sqrt{3},-66i\sqrt{3}, 0,-378\sqrt{3},378i\sqrt{3}, 0,3\sqrt{3},-3i\sqrt{3}).
\eea
Then let us construct the spin 3 representation. The eigenspace of $iM^T$ corresponding to the eigenvalue 3
is 3 dimensional, and in this space the following $u^{(3,3)}$ is orthogonal to $u^{(7,3)}$ and $u^{(5,3)}$:
\bea
u^{(3,3)} & = & (0,0,0, 0,0,0, 0,98,98i, 0,82,-26i, 0,-26,82i, 0,28,28i,
\nn & &
0,-60\sqrt{3},60i\sqrt{3}, 0,12\sqrt{3},-12i\sqrt{3}, 0,27\sqrt{3},9i\sqrt{3},
\nn & &
0,-9\sqrt{3},-27i\sqrt{3}, 0,12\sqrt{3},12i\sqrt{3}, 0,-3\sqrt{3},-3i\sqrt{3}).
\eea
This gives $\ket{---,i}$ in the spin 3 representation:
\beq
\ket{---,i}=u^{(3,3)}_{(Ia)}\ket{I,a,i}.
\eeq
Other states in this representation are obtained by successive action of $ig^+$ on $\ket{---,i}$.
We show the following state in the eigenspace of $iM^T$ corresponding to the eigenvalue 1:
\beq
\ket{-11,i} = -\frac{1}{6}(ig^+)^2\ket{---,i}
 = u^{(3,1)}_{(Ia)}\ket{I,a,i},
\eeq
where
\bea
u^{(3,1)} & = & \frac{1}{2}\cdot(0,0,0, 0,0,0, 0,98,-98i, 0,-62,-118i, 0,118,62i, 0,28,-28i,
\nn & &
0,100\sqrt{3},100i\sqrt{3},0,-20\sqrt{3},-20i\sqrt{3}, 0,3\sqrt{3},-33i\sqrt{3}, 
\nn & &
0,-33\sqrt{3},3i\sqrt{3}, 0,12\sqrt{3},-12i\sqrt{3}, 0,-3\sqrt{3},3i\sqrt{3}).
\eea
Finally, let us construct the spin 1 representation. The eigenspace of $iM^T$ corresponding to the eigenvalue 1
is 4 dimensional, and in this space the following $u^{(1,1)}$ is orthogonal to 
$u^{(7,1)}$, $u^{(5,1)}$ and $u^{(3,1)}$:
\bea
u^{(1,1)} & = & \frac{1}{\sqrt{2}}\cdot(0,0,0, 0,0,0, 0,148,-148i, 0,-112,112i, 0,-112,112i, 0,-112,112i,
\nn & &
0,0,0, 0,0,0, 0,13\sqrt{3},-13i\sqrt{3}, 0,-13\sqrt{3},13i\sqrt{3}, 
\nn & &
0,182\sqrt{3},-182i\sqrt{3}, 0,-13\sqrt{3},13i\sqrt{3}).
\eea
This gives $\ket{-,i}$ in the spin 1 representation:
\beq
\ket{-,i}=u^{(1,1)}_{(Ia)}\ket{I,a,i}.
\eeq
From $\ket{-,i}$ we can construct $\ket{1,i}$ and $\ket{+,i}$.
Then we see that those three states are written in the following compact form ($a=1,2,3$):
\bea
\ket{a,i} & = & 148\ket{3,a,i}-112\left(\ket{4,a,i}+\ket{5,a,i}+\ket{6,a,i}\right) \nn
& & +13\sqrt{3}\left(\ket{9,a,i}-\ket{10,a,i}+14\ket{11,a,i}-\ket{12,a,i}\right).
\eea
Now it is easy to confirm the full transformation property of these states: 
\beq
G^a\ket{b,i}=-if_{abc}\ket{c,i}.
\eeq

Actually we can easily construct a set of states in (adjoint)$\times$(vector) representation of SU(2)$\times$SO(9)
from the singlet state $\ket{S}$:
\beq
f_{abc}\theta_\alpha^b(\gamma^i)_{\alpha\beta}\theta_\beta^c\ket{S},
\eeq
and this should be proportional to $\ket{a,i}$. Indeed, we have confirmed that
\beq
\ket{a,i}=-\frac{351}{4f_{123}}f_{abc}\theta_\alpha^b(\gamma^i)_{\alpha\beta}\theta_\beta^c\ket{S}.
\eeq

Since only $\ket{a,i}$ transforms as adjoint representation of the gauge group,
$\ket{\phi_i^a}$ must be proportional to $\ket{a,i}$,
and now the question is if it satisfies \eqref{firstterm} or not.
To calculate $\gamma^i\theta^a\ket{a,i}$, we first note that $\gamma^i\theta^2\ket{I,2,i}$ and
$\gamma^i\theta^3\ket{I,3,i}$ are given from $\gamma^i\theta^1\ket{I,1,i}$
by the following replacement:
\bea
(\gamma^i\theta^2)_\alpha\ket{I,2,i} & = & (\gamma^i\theta^1)_\alpha\ket{I,1,i}
\Big|_{\ket{*_1}_1\ket{*_2}_2\ket{*_3}_3\rightarrow \ket{*_1}_2\ket{*_2}_3\ket{*_3}_1},
\\
(\gamma^i\theta^3)_\alpha\ket{I,3,i} & = & (\gamma^i\theta^1)_\alpha\ket{I,1,i}
\Big|_{\ket{*_1}_1\ket{*_2}_2\ket{*_3}_3\rightarrow \ket{*_1}_3\ket{*_2}_1\ket{*_3}_2},
\eea
and we need the following Fierz transformations to rewrite $\gamma^i\theta^2\ket{I,2,i}$ and
$\gamma^i\theta^3\ket{I,3,i}$ in the same form as $\gamma^i\theta^1\ket{I,1,i}$:
\bea
(\Gamma_1)_{\alpha\beta}\ket{\delta k}_1\ket{\beta i}_2\ket{\gamma j}_3(\Gamma_2)_{\gamma\delta}
& = & \frac{1}{16}\sum_{n=0}^4\frac{1}{n!}
\nn & & 
\times(\gamma^{i_1\dots i_n})_{\alpha\beta}
\ket{\beta k}_1\ket{\gamma i}_2\ket{\delta j}_3(\Gamma_1^T\gamma^{i_1\dots i_n}\Gamma_2^T)_{\gamma\delta},
\\
(\Gamma_1)_{\alpha\beta}\ket{\gamma j}_1\ket{\delta k}_2\ket{\beta i}_3(\Gamma_2)_{\gamma\delta}
& = & \frac{1}{16}\sum_{n=0}^4\frac{1}{n!}(-1)^{\frac{1}{2}n(n-1)}
\nn & & 
\times(\gamma^{i_1\dots i_n})_{\alpha\beta}
\ket{\beta j}_1\ket{\gamma k}_2\ket{\delta i}_3(\Gamma_2^T\gamma^{i_1\dots i_n}\Gamma_1)_{\gamma\delta}.
\eea
After straightforward calculation using these we obtain
\bea
(\gamma^i\theta^a)_\alpha\ket{3,a,i} & = & 
-\frac{11}{3}\ket{1,\alpha}-\frac{11}{24}\ket{2,\alpha}-\frac{11}{24}\ket{3,\alpha}
\nn & & 
+\frac{11}{24}\ket{5,\alpha}+\frac{11}{24}\ket{6,\alpha}-\frac{11}{48}\ket{8,\alpha}
+\frac{11}{48}\ket{9,\alpha}
\nn & & 
+\frac{11}{144}\ket{12,\alpha}-\frac{11}{144}\ket{13,\alpha},
\\
(\gamma^i\theta^a)_\alpha\ket{4,a,i} & = & 
-\frac{1}{24}\ket{1,\alpha}+\frac{1}{4}\ket{2,\alpha}-\frac{5}{8}\ket{3,\alpha}-\frac{3}{8}\ket{4,\alpha}
\nn & & 
-\frac{7}{12}\ket{5,\alpha}+\frac{13}{24}\ket{6,\alpha}-\frac{7}{48}\ket{7,\alpha}-\frac{1}{24}\ket{8,\alpha}
\nn & & 
-\frac{1}{16}\ket{9,\alpha}+\frac{1}{288}\ket{11,\alpha}+\frac{1}{72}\ket{12,\alpha}+\frac{1}{144}\ket{13,\alpha}
\nn & & 
-\frac{1}{1152}\ket{15,\alpha},
\\
(\gamma^i\theta^a)_\alpha\ket{5,a,i} & = & 
-\frac{1}{24}\ket{1,\alpha}-\frac{5}{8}\ket{2,\alpha}+\frac{1}{4}\ket{3,\alpha}+\frac{3}{8}\ket{4,\alpha}
\nn & & 
+\frac{13}{24}\ket{5,\alpha}-\frac{7}{12}\ket{6,\alpha}+\frac{7}{48}\ket{7,\alpha}+\frac{1}{16}\ket{8,\alpha}
\nn & & 
+\frac{1}{24}\ket{9,\alpha}+\frac{1}{288}\ket{11,\alpha}-\frac{1}{144}\ket{12,\alpha}-\frac{1}{72}\ket{13,\alpha}
\nn & & 
-\frac{1}{1152}\ket{15,\alpha},
\\
(\gamma^i\theta^a)_\alpha\ket{6,a,i} & = & 
-\frac{7}{12}\ket{1,\alpha}+\frac{7}{24}\ket{2,\alpha}+\frac{7}{24}\ket{3,\alpha}+\frac{1}{8}\ket{5,\alpha}
\nn & & 
+\frac{1}{8}\ket{6,\alpha}-\frac{1}{16}\ket{8,\alpha}+\frac{1}{16}\ket{9,\alpha}-\frac{1}{48}\ket{11,\alpha}
\nn & & 
+\frac{1}{144}\ket{12,\alpha}-\frac{1}{144}\ket{13,\alpha}-\frac{1}{576}\ket{15,\alpha},
\\
(\gamma^i\theta^a)_\alpha\ket{9,a,i} & = & \frac{1}{\sqrt{3}}\Big[
-\frac{1}{12}\ket{1,\alpha}+\frac{1}{2}\ket{2,\alpha}-\frac{5}{4}\ket{3,\alpha}+\frac{5}{4}\ket{4,\alpha}
\nn & & 
+\frac{5}{6}\ket{5,\alpha}-\frac{11}{12}\ket{6,\alpha}+\frac{1}{24}\ket{7,\alpha}+\frac{1}{3}\ket{8,\alpha}
\nn & & 
+\frac{1}{6}\ket{10,\alpha}+\frac{1}{8}\ket{9,\alpha}+\frac{1}{48}\ket{11,\alpha}+\frac{1}{18}\ket{12,\alpha}
\nn & &
-\frac{1}{72}\ket{13,\alpha}-\frac{1}{12}\ket{14,\alpha}+\frac{5}{576}\ket{15,\alpha}
\Big], \\
(\gamma^i\theta^a)_\alpha\ket{10,a,i} & = & \frac{1}{\sqrt{3}}\Big[
\frac{1}{12}\ket{1,\alpha}-\frac{1}{2}\ket{3,\alpha}+\frac{5}{4}\ket{2,\alpha}+\frac{5}{4}\ket{4,\alpha}
\nn & & 
-\frac{5}{6}\ket{6,\alpha}+\frac{11}{12}\ket{5,\alpha}+\frac{1}{24}\ket{7,\alpha}+\frac{1}{3}\ket{9,\alpha}
\nn & & 
-\frac{1}{6}\ket{10,\alpha}+\frac{1}{8}\ket{8,\alpha}-\frac{1}{48}\ket{11,\alpha}-\frac{1}{72}\ket{12,\alpha}
\nn & &
+\frac{1}{18}\ket{13,\alpha}+\frac{1}{12}\ket{14,\alpha}-\frac{5}{576}\ket{15,\alpha}
\Big], \\
(\gamma^i\theta^a)_\alpha\ket{11,a,i} & = & \frac{1}{\sqrt{3}}\Big[
\frac{31}{12}\ket{1,\alpha}-\frac{3}{2}\ket{2,\alpha}-\frac{3}{2}\ket{3,\alpha}-\frac{7}{12}\ket{5,\alpha}
\nn & & 
-\frac{7}{12}\ket{6,\alpha}+\frac{5}{24}\ket{8,\alpha}-\frac{1}{6}\ket{10,\alpha}-\frac{5}{24}\ket{9,\alpha}
\nn & & 
+\frac{1}{24}\ket{11,\alpha}-\frac{1}{18}\ket{12,\alpha}+\frac{1}{18}\ket{13,\alpha}+\frac{1}{12}\ket{14,\alpha}
\nn & & 
-\frac{1}{72}\ket{15,\alpha}
\Big], \\
(\gamma^i\theta^a)_\alpha\ket{12,a,i} & = & \frac{1}{\sqrt{3}}\Big[
-\frac{105}{4}\ket{2,\alpha}-\frac{105}{4}\ket{3,\alpha}-\frac{15}{4}\ket{5,\alpha}-\frac{15}{4}\ket{6,\alpha}
\nn & & 
+\frac{15}{8}\ket{8,\alpha}-\frac{15}{8}\ket{9,\alpha}+\ket{11,\alpha}+\frac{3}{8}\ket{12,\alpha}
-\frac{3}{8}\ket{13,\alpha}
\Big],
\eea
where
\bea
\ket{1,\alpha} & = & \ket{\alpha i}_1\ket{\gamma j}_2\ket{\delta j}_3(\gamma^i)_{\gamma\delta},
\\
\ket{2,\alpha} & = & (\gamma^i)_{\alpha\beta}\ket{\beta j}_1\ket{\gamma i}_2\ket{\gamma j}_3,
\\
\ket{3,\alpha} & = & (\gamma^i)_{\alpha\beta}\ket{\beta j}_1\ket{\gamma j}_2\ket{\gamma i}_3,
\\
\ket{4,\alpha} & = & (\gamma^i)_{\alpha\beta}\ket{\beta j}_1\ket{\gamma k}_2\ket{\delta k}_3(\gamma^{ij})_{\gamma\delta},
\\
\ket{5,\alpha} & = & (\gamma^{ij})_{\alpha\beta}\ket{\beta k}_1\ket{\gamma i}_2\ket{\delta k}_3(\gamma^{j})_{\gamma\delta},
\\
\ket{6,\alpha} & = & (\gamma^{ij})_{\alpha\beta}\ket{\beta k}_1\ket{\gamma k}_2\ket{\delta i}_3(\gamma^{j})_{\gamma\delta},
\\
\ket{7,\alpha} & = & (\gamma^{ij})_{\alpha\beta}\ket{\beta k}_1\ket{\gamma l}_2\ket{\delta l}_3(\gamma^{ijk})_{\gamma\delta},
\\
\ket{8,\alpha} & = & (\gamma^{ijk})_{\alpha\beta}\ket{\beta l}_1\ket{\gamma i}_2\ket{\delta l}_3(\gamma^{jk})_{\gamma\delta},
\\
\ket{9,\alpha} & = & (\gamma^{ijk})_{\alpha\beta}\ket{\beta l}_1\ket{\gamma l}_2\ket{\delta i}_3(\gamma^{jk})_{\gamma\delta},
\\
\ket{10,\alpha} & = & (\gamma^{ijk})_{\alpha\beta}\ket{\beta l}_1\ket{\gamma i}_2\ket{\delta j}_3(\gamma^{kl})_{\gamma\delta},
\\
\ket{11,\alpha} & = & (\gamma^{ijk})_{\alpha\beta}\ket{\beta l}_1\ket{\gamma m}_2\ket{\delta m}_3(\gamma^{ijkl})_{\gamma\delta},
\\
\ket{12,\alpha} & = & (\gamma^{ijkl})_{\alpha\beta}\ket{\beta m}_1\ket{\gamma i}_2\ket{\delta m}_3(\gamma^{jkl})_{\gamma\delta},
\\
\ket{13,\alpha} & = & (\gamma^{ijkl})_{\alpha\beta}\ket{\beta m}_1\ket{\gamma m}_2\ket{\delta i}_3(\gamma^{jkl})_{\gamma\delta},
\\
\ket{14,\alpha} & = & (\gamma^{ijkl})_{\alpha\beta}\ket{\beta m}_1\ket{\gamma i}_2\ket{\delta j}_3(\gamma^{klm})_{\gamma\delta},
\\
\ket{15,\alpha} & = & (\gamma^{ijkl})_{\alpha\beta}\ket{\beta m}_1\ket{\gamma n}_2\ket{\delta n}_3(\gamma^{ijklm})_{\gamma\delta}.
\eea
Therefore,
\bea
(\gamma^i\theta^a)_\alpha\ket{a,i} & = & -13\ket{8,\alpha}+13\ket{9,\alpha}-26\ket{10,\alpha}
-\frac{239}{72}\ket{11,\alpha} \nn
& & -\frac{13}{3}\ket{12,\alpha}+\frac{13}{3}\ket{13,\alpha}+13\ket{14,\alpha}
-\frac{551}{288}\ket{15,\alpha}.
\label{githaai}
\eea
In this calculation we did not take $\gamma^{12\dots 9}=1$ into account. 
If we use it we see that the above 15 states $\ket{I,\alpha}$ are not independent.
First note that the following hold from $\gamma^{12\dots 9}=1$:
\beq
\gamma^{i_1\dots i_n}=\frac{(-1)^{\frac{1}{2}n(n-1)}}{(9-n)!}\epsilon^{i_1\dots i_9}\gamma^{i_{n+1}\dots i_9,}
\eeq
and
\beq
(\gamma^{i_1\dots i_n})_{\alpha\beta}(\gamma^{i_1\dots i_nj_1\dots j_m})_{\gamma\delta}
=(-1)^{\frac{1}{2}m(m+2n-1)}\frac{n!}{(9-n-m)!}(\gamma^{j_1\dots j_mk_1\dots k_{9-n-m}})_{\alpha\beta}
(\gamma^{k_1\dots k_{9-n-m}})_{\gamma\delta}.
\label{epsgamma}
\eeq
With \eqref{epsgamma} and the Rarita-Schwinger condition, we can show the following:
\bea
\ket{15,\alpha} & = & 
(\gamma^{ijklm})_{\alpha\beta}\ket{\beta m}_1\ket{\gamma n}_2\ket{\delta n}_3(\gamma^{ijkl})_{\gamma\delta}
\nn & = & -4(\gamma^{ijk})_{\alpha\beta}\ket{\beta m}_1\ket{\gamma n}_2\ket{\delta n}_3(\gamma^{ijkm})_{\gamma\delta}
\nn & = & -4\ket{11,\alpha}.
\label{notid1}
\eea
Furthermore, from the Rarita-Schwinger condition, $0=(\gamma^{ijkl}\gamma^m)_{\alpha\beta}\ket{\beta m}_1
\ket{\gamma i}_2\ket{\delta j}_3(\gamma^{kl})_{\gamma\delta}$, and by expanding the product of the gamma matrices,
\bea
0 & = & (\gamma^{ijklm})_{\alpha\beta}\ket{\beta m}_1\ket{\gamma i}_2\ket{\delta j}_3(\gamma^{kl})_{\gamma\delta}
+2(\gamma^{ijk})_{\alpha\beta}\ket{\beta m}_1\ket{\gamma i}_2\ket{\delta j}_3(\gamma^{km})_{\gamma\delta}
\nn & &
+(\gamma^{ikl})_{\alpha\beta}\ket{\beta j}_1\ket{\gamma i}_2\ket{\delta j}_3(\gamma^{kl})_{\gamma\delta}
-(\gamma^{jkl})_{\alpha\beta}\ket{\beta i}_1\ket{\gamma i}_2\ket{\delta j}_3(\gamma^{kl})_{\gamma\delta}
\nn & = & 
\ket{8,\alpha}-\ket{9,\alpha}+2\ket{10,\alpha}
+(\gamma^{ijklm})_{\alpha\beta}\ket{\beta m}_1\ket{\gamma i}_2\ket{\delta j}_3(\gamma^{kl})_{\gamma\delta}.
\eea
Applying \eqref{epsgamma} to the last term of the above and expanding the product of gamma matrices, we obtain
\beq
0=\ket{8,\alpha}-\ket{9,\alpha}+2\ket{10,\alpha}+\frac{1}{3}\ket{12,\alpha}-\frac{1}{3}\ket{13,\alpha}
-\ket{14,\alpha}+\frac{1}{12}\ket{15,\alpha}.
\label{notid2}
\eeq
Similarly, from $0=(\gamma^{ijklm}\gamma^n)_{\alpha\beta}\ket{\beta n}_1
\ket{\gamma i}_2\ket{\delta j}_3(\gamma^{klm})_{\gamma\delta}$, we obtain
\beq
0=3\ket{8,\alpha}-3\ket{9,\alpha}+6\ket{10,\alpha}-\ket{11,\alpha}-\ket{12,\alpha}+\ket{13,\alpha}
+3\ket{14,\alpha}.
\label{notid3}
\eeq
Using \eqref{notid1}, \eqref{notid2} and \eqref{notid3}, 
we can see that the right hand side of \eqref{githaai} is in fact zero:
\beq
\gamma^i\theta^a\ket{a,i}=0.
\eeq
Therefore $\ket{\phi_i^a}$ can be proportional to $\ket{a,i}$ with a nonzero coefficient.

%%%%%%%%%%%%%%%%%%%%%%%%%%%%%%%%%%%%%%%%%%%%%%%%%%%%%%%%%%%%%%%%%%%
\section{Discussion}

We have found that 36 vector states are classified into spin 7, spin 5, spin 3 and spin 1 
representation of SU(2). In addition, we have shown that
the linear term in $X_i^a$ in Taylor expansion of the zero energy wavefunction around the origin
is proportional to the spin 1 representation.

Clearly brute-force calculation as the one we have made in this paper is too inefficient 
to construct states in higher representations of SO(9), and we have to invent more 
useful method to analyze them.
We have found that $\ket{a,i}$ is given by acting $\theta_\alpha^a$ on the singlet $\ket{S}$.
Other states may be constructed similarly. For example, the traceless part of the following state:
\beq
\sum_{(a_1a_2a_3)}
[f_{a_1b_1c_1}\theta^{b_1}_{\alpha_1}(\gamma^i)_{\alpha_1\beta_1}\theta^{c_1}_{\beta_1}]
[f_{a_2b_2c_2}\theta^{b_2}_{\alpha_2}(\gamma^j)_{\alpha_2\beta_2}\theta^{c_2}_{\beta_2}]
[f_{a_3b_3c_3}\theta^{b_3}_{\alpha_3}(\gamma^j)_{\alpha_3\beta_3}\theta^{c_3}_{\beta_3}]\ket{S},
\eeq
where $\sum_{(a_1a_2a_3)}$ is summation over permutations of $a_1a_2a_3$,
must be proportional to $\ket{a_1a_2a_3,i}$. The question is 
if the proportionality constant is zero or not.
We did not check it, and similar problems arise for other representations.
In general it is an interesting problem if there are
states which cannot be constructed from $\ket{S}$ or not.

Before constructing each states explicitly it is desirable to know multiplicities of representations
in the $2^{8(N^2-1)}$-dimensional vector space\cite{micta}. Related counting has been made in \cite{trze07}.

%%%%%%%%%%%%%%%%%%%%%%%%%%%%%%%%%%%%%%%%%%%%%%%%%%%%%%%%%%%%%%
\vs{.5cm}
\noindent
{\large\bf Acknowledgments}\\[.2cm]
I would like to thank M.~Hynek and M.~Trzetrzelewski for correspondence.
%%%%%%%%%%%%%%%%%%%%%%%%%%%%%%%%%%%%%%%%%%%%%%%%%%%%%%%%%%%%%%

%%%%%%%%%%%% References %%%%%%%%%%%%%%%%%%%%%%%%%
\newcommand{\J}[4]{{\sl #1} {\bf #2} (#3) #4}
\newcommand{\andJ}[3]{{\bf #1} (#2) #3}
\newcommand{\AP}{Ann.\ Phys.\ (N.Y.)}
\newcommand{\MPL}{Mod.\ Phys.\ Lett.}
\newcommand{\NP}{Nucl.\ Phys.}
\newcommand{\PL}{Phys.\ Lett.}
\newcommand{\PR}{Phys.\ Rev.}
\newcommand{\PRL}{Phys.\ Rev.\ Lett.}
\newcommand{\PTP}{Prog.\ Theor.\ Phys.}
\newcommand{\hepth}[1]{{\tt hep-th/#1}}
\newcommand{\arxivhep}[1]{{\tt arXiv.org:#1 [hep-th]}}
%%%%%%%%%%%%%%%%%%%%%%%%%%%%%%%%%%%%%%%%%%%%%%%%

%%%%%%%%%%%%%%%%%%%%%%%%%%%%%%%%%%%%%%%%%%%%%%%%%%%%%%%%%%%%%%%%%%%
\end{document}